\documentclass[10pt, conference, letterpaper]{IEEEtran}
\IEEEoverridecommandlockouts
\usepackage{cite}
\usepackage{amsmath,amssymb,amsfonts}
\usepackage{algorithmic}
\usepackage{graphicx}
\usepackage{textcomp}
\usepackage{xcolor}
\usepackage{CJKutf8}
\usepackage{enumerate}
\usepackage{bicaption}
\usepackage{subfiles}
\usepackage{bm}
\usepackage{float}
\usepackage{CJKutf8}
\usepackage{subfigure}
\usepackage{multirow}
\def\BibTeX{{\rm B\kern-.05em{\sc i\kern-.025em b}\kern-.08em
T\kern-.1667em\lower.7ex\hbox{E}\kern-.125emX}}
\begin{document}
\begin{CJK}{UTF8}{gbsn}

\title{Electromagnetic Signal Modulation Recognition based on Subgraph Embedding Learning}
\author{
    Bojun Zhang\IEEEauthorrefmark{1} % 作者名和标记
    \thanks{
    \IEEEauthorrefmark{1}Bojun Zhang is the corresponding author. Tianjin University, China Tianjin 300000, China. 
    Email: ewwllvraier@126.com
}
}
\maketitle
\begin{abstract}
Automatic Modulation Recognition (AMR) detects modulation schemes of received signals for further processing of signals without any priori information, which is critically important for civil spectrum regulation, information countermeasures, and communication security.
%
%Recent breakthroughs in deep learning (DL) have laid the foundation for the development of high-performance DL-AMR algorithms for communication systems. 
%
Due to the powerful feature extraction and classification capabilities of Deep Learning (DL), DL-based AMR algorithms have achieved excellent performance gains compared with traditional modulation detection algorithms. 
However, all existing DL-based AMR algorithms, to the best of our knowledge, are designed for specific channels and systems, because data dimension of the used training dataset is fixed.
To this end, we takes the first step to propose a Subgraph Embedding Learning (SEL) structure to address the classical AMR problem, and the proposed algorithm is called SEL-AMR. 
Our algorithm treats the communication system as a subgraph and uses the relationship between samples to smooth the effects brought by noise and different channels to extract robust features. 
Thus, the proposed SEL-AMR algorithm can adapt to any dynamic channels and systems. 
We use 5 public real datasets and a small amount of simulation data to evaluate our SEL-AMR algorithm. 
Experimental results reveal that SEL-AMR can well adapt to different channels and systems, and always outperforms the state-of-the-art algorithms by improving up to 20\% macro-average recognition precision and 30\% recognition accuracy. 

\end{abstract}


\section{Introduction}
% \subfile{section/intro.tex}
\subsection{Background and Motivation}
\begin{figure}[htbp]
	\centering
	\includegraphics[width=\linewidth,scale=1.50]{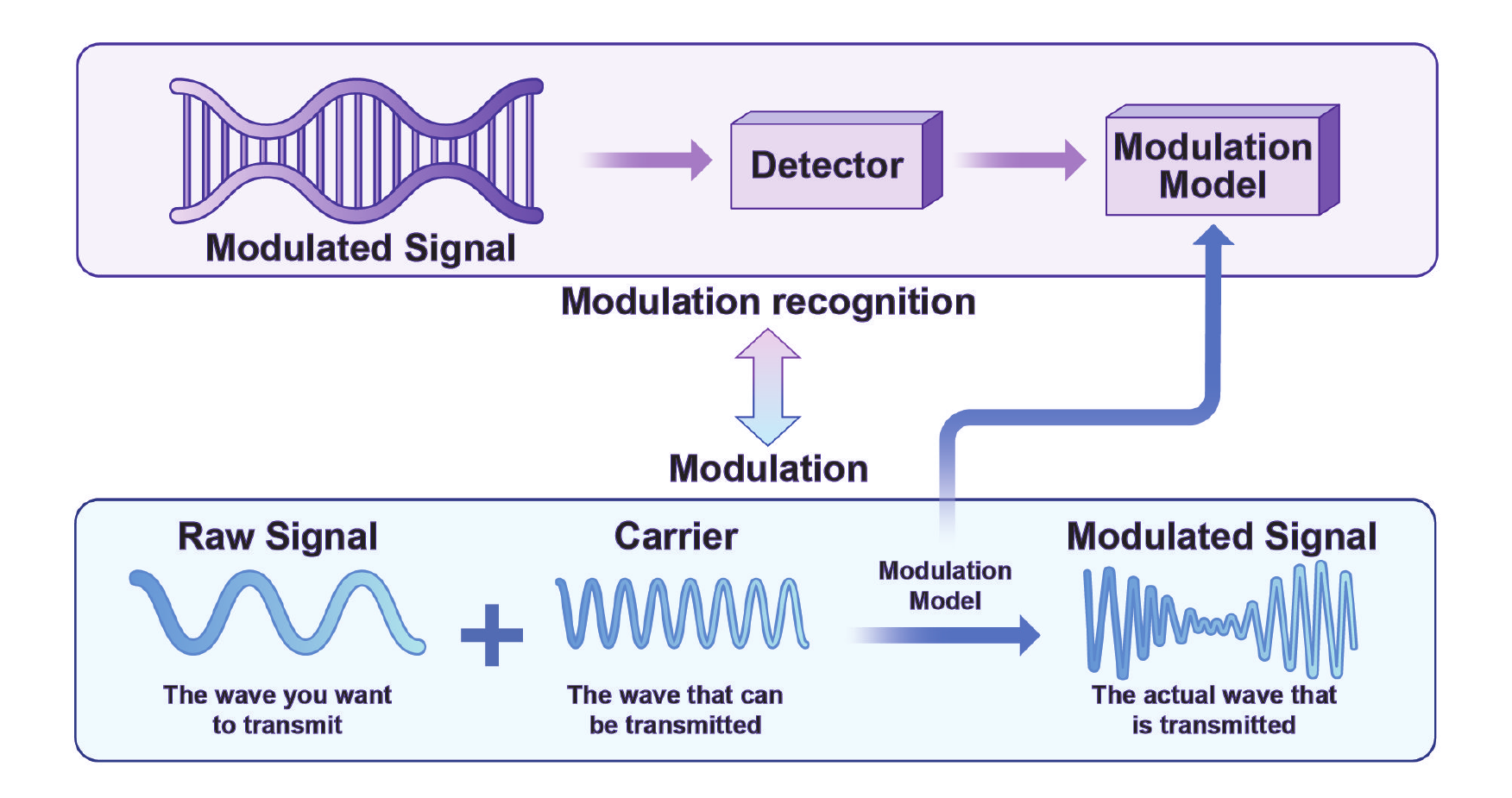}
	\caption{Signal Modulation Recognition}
	\label{SMR}
        \vspace{-0.60cm}
\end{figure}

With the rapid development of wireless communication technology, the use of wireless technology for steganography has gradually become
the mainstream of technical steganography, which brings great challenges to the whole environment security protection. 
As illustrated in Fig. \ref{SMR}, signal modulation recognition, as the intermediate process of signal detection and demodulation, is widely used in the field of signal detection and
recognition, which can automatically judge and identify the modulation mode of unknown signals and then provide modulation information
for the correct demodulation of illegal signals. Modulation recognition technology has a wide range of applications in military electronic
countermeasures, civilian spectrum monitoring and management, x-information countermeasures, communication security, software radio and
cognitive radio and has become an important research topic and hot topic at present.
% \vspace{-0.5cm}
\subsection{Limitations of Prior Arts}
Modulation recognition is mainly divided into automatic modulation recognition and manual modulation recognition. 
Among them, manual modulation identification mainly relies on experienced operators to carry out by analyzing the signal waveform and spectrum changes displayed by the instrument. 
This requires a lot of experience and is inefficient. With the development of technology, automatic modulation identification has replaced most of the manual identification needs.
Automatic modulation recognition algorithms (AMRs) can be broadly classified into two categories: likelihood estimation-based and machine learning-based
approaches\cite{zhang2022deep}. The automatic modulation recognition algorithm based on likelihood estimation is actually a multicomposite hypothesis testing
problem. According to the different models established for unknown quantities, AMR algorithms based on likelihood estimation can be broadly
classified into three types: mean likelihood ratio test\cite{sills1999maximum}, generalized likelihood ratio test\cite{lay1995modulation}
and mixed likelihood ratio test\cite{hong2000bpsk}. With the development of machine learning, increasingly more work is being done using machine
learning algorithms for recognition\cite{yuan2004modulation}\cite{park2008automatic}. 
% Jiang \emph{\emph{et al}}\cite{yuan2004modulation}. use decision trees to construct prediction models.
% Park \emph{\emph{et al}}\cite{park2008automatic}. used SVM for automatic modulation recognition. 
Deep learning has already made breakthrough progress in many challenging applications that are difficult for traditional algorithms to achieve.
Some pioneering work\cite{chen2022emd}\cite{zhang2022variable}\cite{wang2020deep} has adopted novel deep learning-based approaches, which often outperform conventional algorithms for modulation recognition tasks.
%
% Deep learning (DL) has made breakthroughs in a
% range of challenging applications that are difficult to achieve with traditional algorithms, with some of this work proposing groundbreaking
% deep learning-based approaches that often outperform traditional modulation recognition algorithms. Tao \emph{\emph{\emph{et al}}}. \cite{chen2022emd} used the Hilbert transform
% combined with deep learning to extract deeper features. 
% %
% Zhang \emph{\emph{\emph{et al}}}. \cite{zhang2022variable} used domain adversarial neural networks to
% generate domain-invariant fingerprint features. 
% %
% Wang \emph{\emph{\emph{et al}}}. \cite{wang2020deep} used CNN for modulation recognition of MIMO systems.
Despite the great potential of deep learning-based AMR algorithms, most existing AMR algorithms are designed for fixed scenarios and
specific systems and cannot be extended to scenarios where the environment and channel conditions differ significantly from the training
data. To address the above issues, we propose a new accurate and robust algorithm to solve the classical electromagnetic signal modulation
recognition problem.
\subsection{Nutshell Algorithm}
In this paper, we consider the signal data samples as a large graph and each signal system as a subgraph of the large graph. We first
align, complement, and normalize each signal data by preprocessing, and then we use our constructed MIMO embedding module to use the idea
of subgraph embedding to aggregate the features of each signal subgraph using a graph isomorphism network and use set2set to map the
graph features in non-Euclidean space to Euclidean space to obtain the embedding vector of the subgraph. Next, the topology of the large
graph composed of the sample data will be constructed using the KNN cluster algorithm. Finally, the GAT-LPA algorithm is used to smooth
out the effects of noise and different channel conditions by using the relationship between samples to obtain a robust classification
result.
\subsection{Challenges and Solutions}
\emph{The first challenge of this paper is to make the model applicable to different signal systems}. 
The difficulty of this challenge lies in the fact that there are obvious differences in the structure and characteristics of different signal systems, and the contradiction between diversity and normalization needs to be dealt with in order to find their common characteristics to build a unified representation model. 
The diversity of features of different types of signaling systems adds difficulties to modeling, while mapping them into a unified space requires abstracting their normality. 
Dealing with this conflict between diversity and normalization is the main difficulty of this challenge.
Our solution is to consider the signal system as a bipartite graph, use a graph isomorphism network, a model applicable to the changeable adjacency matrix, to extract the features of the bipartite graph and use set2set to construct a non-Euclidean space to Euclidean space and order-independent mapping to obtain the embedding vector of each signal system to map different types of signal systems to a uniform space.

\emph{The second challenge is to construct the topology between data in signal sample data that do not show the graph structure}. 
Since the original data do not have a graph network structure, to construct the connectivity between nodes, it is necessary to design reasonable clustering and embedding algorithms, and the lack of direct explicit graph structure increases the difficulty of constructing the topology, and reasonable design methods are crucial for obtaining the connectivity between the sample data.
We use the KNN graph approach to construct a learnable adjacency matrix between signal samples by means of clustering.

\emph{The third challenge is how to ensure the accuracy of the model and the ability to generalize to unlabeled data when facing a large
amount of unlabeled data. }
Due to the existence of massive unlabeled data seriously affects the generalization of the model, to ensure the prediction effect needs to deal with the relationship between labeled and unlabeled data, as well as the self-learning ability of the model, which puts forward a very high demand on the model's generalization ability and needs to consider the relationship between the labeled and unlabeled data modeling as well as the self-learning at the same time, which increases the complexity of the problem.
We use the LPA algorithm to assist GAT in prediction and training the label transfer matrix required for LPA.
through GAT and then use LPA to diffuse the labels of the labeled data to the unlabeled data through the trained transfer matrix
and the topology between the data.
\subsection{Novelty and Advantages over Prior Arts}
The innovation of this paper lies in the following three aspects:
\boldmath{(i)} In the field of automatic electromagnetic signal modulation, we present for the first time an adaptive signal inference algorithm that can explicitly represent the hidden relationships between multiple signal data and is applicable to any signal system.
\boldmath{(ii)} We use the KNN graph approach to construct graphs in data where there is no display graph structure, such as signal samples, and we use the spatial partitioning algorithm of ball trees to reduce the time complexity required for clustering.
\boldmath{(iii)} We use the inception structure and residual connections to build our model architecture, thus avoiding to some extent the possible over-smoothing phenomenon of graph neural networks.

% The main advantage of the proposed algorithm is that we can achieve more accurate and robust electromagnetic signal modulation recognition.
% results by the subgraph embedding technique. To validate the performance of the algorithm in this paper, we verified it on several datasets.
The main advantage of the proposed algorithm is that it can achieve more accurate and robust electromagnetic signal modulation recognition through the subgraph embedding technique. To validate the performance of this algorithm, we validate it on several datasets.
Our algorithm provides better accuracy and robustness than the state-of-the-art solutions.

The remainder of this paper is organized as follows.
%
% Section \ref{Preliminary Knowledge} introduce the preparatory knowledge needed to understand this paper. 
%
Section \ref{algorithm OVERVIEW} describes the details of the proposed algorithm. 
We evaluate the performance of our algorithm in Section \ref{IMPLEMENTATION AND EVALUATION}. 
Section \ref{RELATED WORK} discusses related work. 
Section \ref{CONCLUSION} concludes this~paper.

\section{ALGORITHM OVERVIEW}\label{algorithm OVERVIEW}
% \subfile{section/sys.tex}
This section first describes an overview of our proposed algorithm, followed by details of its four main components.
\subsection{Algorithm Overview}
\begin{figure*}[!htp]
    \centering
    \includegraphics[width=0.9\textwidth]{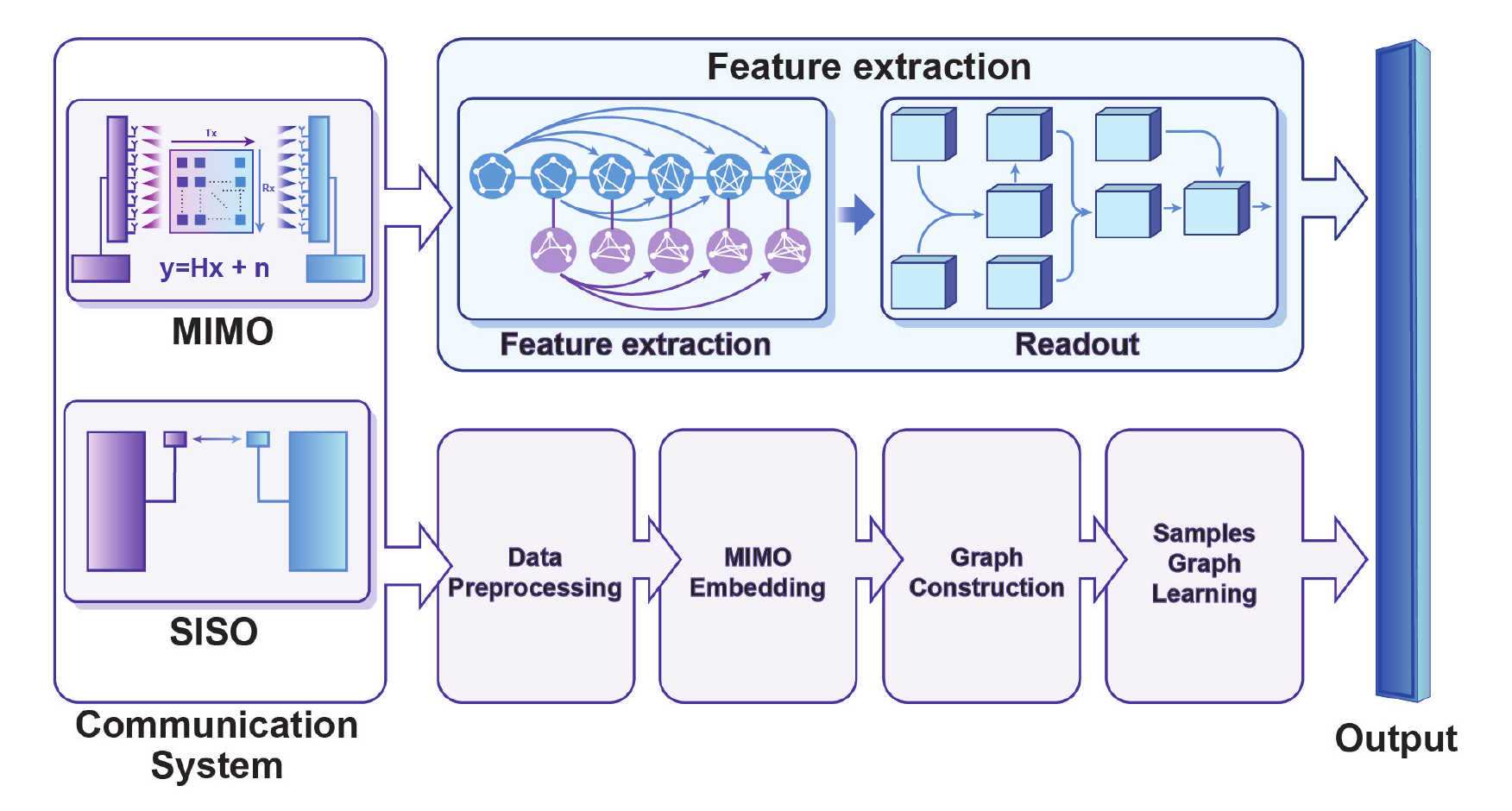}
\caption{Overview of our algorithm}
    \label{fig:frameworkoverview}
    \vspace{-0.5cm}
\end{figure*}
As illustrated in Fig. \ref{fig:frameworkoverview}, the algorithm proposed in this paper can be divided into four parts.
\boldmath{(i) Data Preprocessing: } First, the original electromagnetic signal data are subjected to interpolation operations,
data normalization, mean filtering and wavelet transform. \boldmath{(ii) MIMO Embedding: } We consider all signal samples as a large graph,
and the different communication systems corresponding to each sample as subgraphs on the large graph. For the subgraphs corresponding to
different communication systems, we propose an embedding mechanism that can represent different communication systems such as MIMO, SISO,
etc. as vectors embedded in a Euclidean space. \boldmath{(iii) Graph Construction: } We use the K-nearest neighbor clustering algorithm to construct
learnable topological relationships between different signal samples, denoted as the adjacency matrix of the graph. \boldmath{(iv) Sample Graph Learning: }
Finally, we view the AMR aliasing problem as a semi-supervised node classification problem on the constructed large graph. We use the GAT-LPA
mechanism to extract robust features based on the topology of the data, and a label propagation algorithm to spread the labels of the labeled
data to the unlabeled data based on the relationships between the data to perform signal modulation recognition. Next, we will describe
the details of the above parts.
\subsection{Data Preprocessing}
We hope that our proposed algorithm can be applied to scenarios where multiple communication systems exist, such as SISO systems and
MIMO systems with different numbers of receiving and transmitting antennas. Therefore, we uniformly use IQ data as the input data of
the algorithm. The dimension of the IQ data is $(N_{R}+N_{T})*L*2$. $N_{T}$ is the number of transmitting antennas, $N_{R}$ is the number of
receiving antennas. $L$ is the length of time sampling. We can see that if it is a SISO system, then the data dimension of the system
is $(1+1)*L*2$, which is a special case of a MIMO system.

We first interpolate the IQ data sampled from each antenna separately to ensure that all data are sampled with the same length. After
that the IQ data on the antennas in each sample are deflated to between 0 and 1 using maximum-minimum normalization. Finally, a mean
filtering algorithm is used to perform basic denoising of the data. Finally, the wavelet transform is used to remove outliers from
the data.
\subsection{MIMO Embedding}
To make our algorithm applicable to various signaling systems, we need to construct a mapping of different dimensional data
to the fixed-length data. Our basic idea is to view all samples in the dataset as a large graph, and the signal system corresponding to
each sample as a subgraph.

We first construct the subgraph structure of the signal system corresponding to each sample. Specifically, in the signal system, we
consider each antenna as a node, the channel matrix between different antennas as edges on the graph, and the IQ data on each antenna
as features of the node. We construct the set of nodes $V_T$ consisting of transmit antennas and the set of nodes $V_R$ consisting of receive
antennas. It can be seen that $V_T \cap V_R = \varnothing$. No two nodes of the set of nodes of the transmitting antenna have edges
connected to each other, while no two nodes of the set of nodes of the receiving antenna have edges connected to each other. Since we
use the channel matrix as the edges between nodes, each transmitting antenna node has a connected edge to all receiving antenna nodes,
Similarly, each receiving antenna node has a connected edge to all transmitting antenna nodes. This shows that we map the MIMO system
into a bipartite graph consisting of transmit antenna nodes and output antenna nodes. The SISO system is a special case of the MIMO system,
which is equivalent to a bipartite graph with two nodes. We use the I-component and Q-component of the IQ data as the nodal features
of the nodes on the graph, respectively. At this point, we mapped the signal system into two bipartite graphs with equal adjacency
matrices but different node features.

If we encounter some cases where we need to perform blind estimation of the signal, that is, to estimate the modulation of the signal when the number of transmitting antennas is unknown. We can use the spatial spectral decomposition technique\cite{roy1989esprit} to estimate the number of transmitting antennas. First we can get the number of eigenvalues or singular values by performing eigenvalue decomposition or singular value decomposition on the autocorrelation matrix of the received signal. The number of these eigenvalues or singular values is associated with the number of transmitting antennas. If there are multiple independent transmitting antennas in the system, then the number of eigenvalues or singular values will be greater than or equal to the actual number of transmitting antennas. After that, the spatial spectral decomposition of the received signal can be performed to obtain the power distribution of the signal in different directions. Based on the characteristics of the power distribution, the number of paths through which the signal passes can be inferred, and thus the number of transmitting antennas can be estimated. Finally, after we get the number of transmit antennas we can construct the bipartite graph of the MIMO system according to the procedure introduced before. We use an initial random trainable one-dimensional tensor as the node characteristics of the transmit antenna nodes.
\begin{figure}[htbp]
        \vspace{-0.5cm}
	\centering
	\includegraphics[width=\linewidth,scale=1.50]{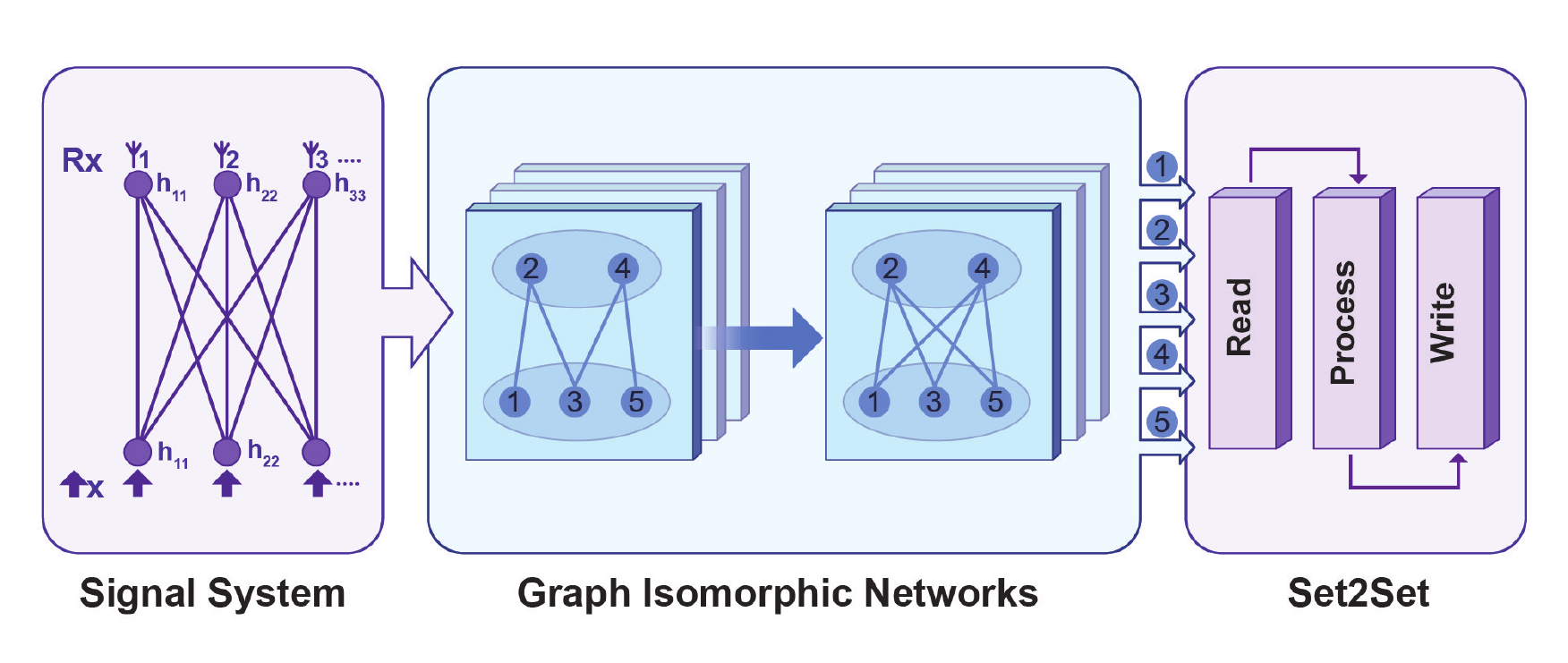}
\caption{MIMO Embedding}
	\label{ME}
        \vspace{-0.3cm}
\end{figure}
As illustrated in Fig. \ref{ME}, after obtaining the graph structure, we first need to extract the features on that graph. Since our graph structure is constantly changing,
A transductive algorithm such as GCN\cite{welling2016semi} is not applicable. Therefore, we choose a graph isomorphic neural network (GIN)\cite{xu2018powerful}
to perform feature extraction on the graph.

The WL test is the upper bound for the GNN to judge graph isomorphism and the upper session of the GNN's ability to extract features. Hence, we design
the graph neural network structure to approximate the WL test as much as possible. the WL test makes a single-shot of a multiset (a
collection of neighbors of a node) one hash function at a time, while a traditional inductive algorithm such as GraphSage\cite{hamilton2017inductive} uses a variety
of functions when aggregating neighbor information, but it is obviously not a single shot. Therefore, we first need to construct the
one-shot aggregation function when aggregating neighbor node features in graph neural networks. If we assume that the set of neighbors
of a node is a countable set and that the set has a finite number of elements, then it is clear that there exists a function $h$, so that
$h$ the single shot can be expressed as a set of single shot functions $f$.
\begin{equation}
  h(X)=\sum_{x \in X} f(x)
\end{equation}
We need an $n$-dimensional vector to store the function $h$, where $n$ is the upper bound on the number of elements of the set, and the value field
of each bit is also $n$, store the number of occurrences of each element, and then we will represent the vector as an $n$-bit $n$-decimal number,
that indeed it can be written as the above representation, where $h(X)$ represents the value of this $n$-decimal number, and $f(x)$ represents the value represented by each bit.

With the above description, we can distinguish the different neighbor structures of each node, but we aim to distinguish the node-centered
Ego-Graph of each node. hence, we also need to consider the node itself, and then we need a single projection function.
\begin{equation}
  h(c, X)=(1+\epsilon) f(c)+\sum_{x \in X} f(x)
\end{equation}
where $c$ denotes the node's own feature y, which is the IQ component, $X$ denotes the set of features of node V's neighbors, $\epsilon$ is an arbitrary
irrational number, but $x,c$ are rational numbers, and $f$ is also a rational function.

Since single-shot functions nested in single-shot functions are obviously also single-shot functions, a final single-shot function $\phi$
is defined again to add learnable parameters to the above equation.
\begin{equation}
  h(c, X)=\phi\left((1+\epsilon) f(c)+\sum_{x \in X} f(x)\right)
\end{equation}
Using the powerful fitting ability of MLP to learn the $\phi$ function and $f$ function, and making $M L P=f^{(l)} \circ \phi^{(l-1)}$,
the final equation of GIN can be obtained.
{\small
\begin{equation}
  h_{k+1}(u)=MLP\left((1+\epsilon) h_k(u)+\sum_{(u, v)\in E}h_k(v)\right)
\end{equation}
}
The above shows our process of extracting the node features on the graph, and we can see that our process does not depend on the adjacency
matrix of the graph. hence, it is suitable for scenarios where the structure of the graph changes, while the propagation process and aggregation
function in the graph neural network both use single-shot functions, so it is closer to the upper bound of the graph neural network's
ability to extract features than the traditional algorithm. Therefore, we use two layers of GIN as the feature extraction layer in our graph
neural network framework.

After we obtain the node features of the graph, we need to extract the features of the whole graph by the node features of the graph and
the topology of the graph. Since we want the features of the graph to be independent of the order of the nodes, instead of using the
traditional sequence generation algorithm, we use set2set\cite{vinyals2015order}, a set-generating set algorithm, as the readout layer in the graph neural
network.
\begin{equation}
  \begin{aligned}
  m_i = MLP(x_i) \\
  q_t  =\operatorname{LSTM}\left(q_{t-1}^*\right) \\
  e_{i, t}  =f\left(m_i, q_t\right) \\
  a_{i, t}  =\frac{\exp \left(e_{i, t}\right)}{\sum_j \exp \left(e_{j, t}\right)} \\
  r_t  =\sum_i a_{i, t} m_i \\
  q_t^*  =\left[q_t r_t\right]
  \end{aligned}
\end{equation}
For all samples $x_i$, a small neural network is used for embedding and encoding as a memory vector $m_i$, where the read block is to feed the
node features after two layers of GIN into a feedforward neural network to embed the input to obtain the corresponding memory vector.
Afterward, our process block uses the LSTM\cite{hochreiter1997long} controller, thus replacing the traditional sequence generation model
to use LSTM directly for generation. Specifically, we use LSTM without input and output, continuously read in the memory vector mi,
repeatedly update the state, repeatedly update T steps, use attention mechanism to describe the previous part, the output of this part,
the hidden state vector of LSTM has sequence invariance for the input data (i.e., the input mi, mj sequential transformation, the hidden
state vector obtained is invariant). The finalized writer block is an LSTM pointer network, whose input is the hidden state output from
the Process part, and the memory vector output from the read part, producing the final output. We use the attention mechanism to obtain
the attention coefficient by combining the output of the encoder, i.e., the hidden state vector from the previous step, and the output of
decoder at the current moment. Here, since the original attention mechanism is used to output the result, an additional attention step
(called glimpse) is performed before the pointer output in this step compared to the original pointer network, and an additional attention
mechanism is added before the pointer part, hence the name glimpse mechanism. Afterward, the probability distribution is obtained based
on the attention coefficient, and the input position with the highest probability is taken as the current output using softmax to learn
the weight relation a. Then, the final graph feature vector is obtained based on weight relation a and implicit state e.
Finally, we obtain the final MIMO embedding vector by weighted fusion of the graph embedding vector obtained from the graph composed
of the I component and the graph embedding vector obtained from the graph composed of the Q component through the channel attention
mechanism\cite{hu2018squeeze}. From this, we construct a mapping of the transformed non-Euclidean graph of the signal system to the embedding
vector in the Euclidean space.

After we obtain the embedding vectors corresponding to the subgraphs of the signal system, we want to use the relationships between
signal samples in complex scenarios to smooth out the effects of data noise and different channel conditions\cite{zhang2022deep}. We use a graph structure
to represent the relationships between signal samples. However, the original sample data do not have a displayed topology between them,
Therefore, we use KNN graphs\cite{dong2011efficient} to construct the topology between the data.
\subsection{Graph Construction}
\begin{figure}[htbp]
	\centering
	\includegraphics[width=\linewidth,scale=1.00]{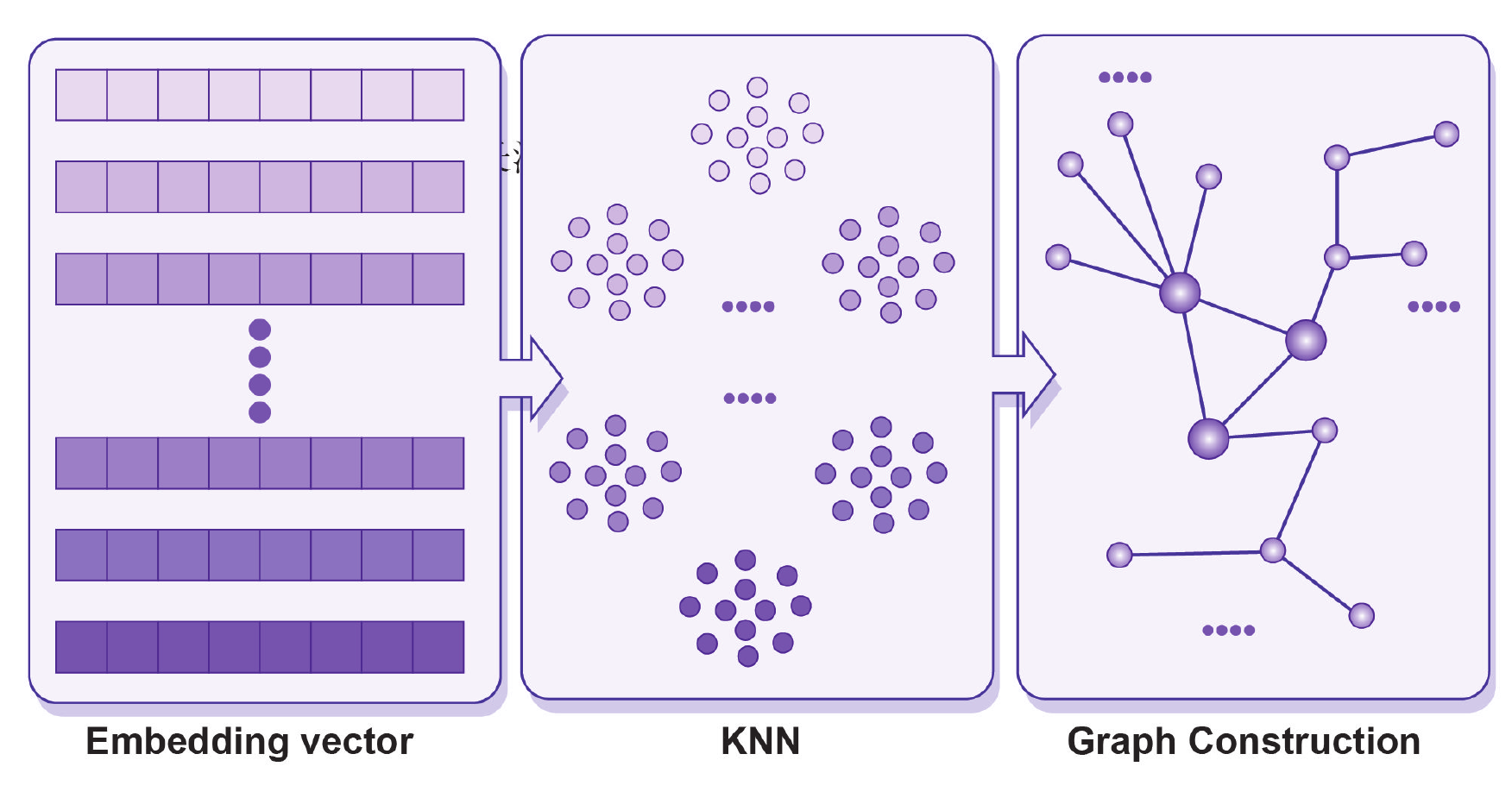}
\caption{KNN Graph}
	\label{knn}
        \vspace{-0.2cm}
\end{figure}
As illustrated in Fig. \ref{knn}, the KNN graph is a step of the composition process added to the classical KNN (k-nearest neighbor) algorithm\cite{abeywickrama2015k}.
Suppose there are N nodes in the space, and for node $N_i$, find the k nearest neighbors $N_1,N_2,⋯,N_k$ by some distance metric (Euclidean
distance, the edit distance, in this paper we use the Euclidean distance. and then connect $N_i$ with these k neighbors to form k directed
edges, respectively. This is done for all nodes in the space, and finally, the KNN graph is obtained.

Since we need to perform spatial partitioning on the entire signal data sample and the signal is embedded in a high-dimensional space after
MIMO embedding, which leads to the traditional spatial partitioning algorithm\cite{bentley1975multidimensional} with time complexity
of $o(n^2)$ and low efficiency in high dimensions. For this reason, we use the ball tree algorithm\cite{pedregosa2011scikit}. The ball
tree divides the data into a series of nested hyperspheres. This makes tree construction more expensive than traditional algorithms such
as KD trees but leads to a data structure that is very efficient for highly structured, even in very high dimensions. The sphere tree
recursively divides the data into nodes defined by the center of mass and radius such that each point in the node lies within and defined
hypersphere, reducing the number of candidate points for neighbor search by using triangular inequalities. Using this setup, a single
distance calculation between the test point and the center of mass is sufficient to determine the upper and lower bounds on the distance
between all points within the node. Due to the spherical geometric properties of the spherical tree nodes, the time complexity of the
algorithm remains $o(nlogn)$, although the actual performance is highly dependent on the structure of the training data, it can represent
a KD tree in high dimensions. From this, we construct the topology between the sample data.
\subsection{Sample Graph Learning}
After obtaining the topology between the data, we transform the original electromagnetic signal automatic modulation recognition (AMR)
problem into a semi-supervised node classification problem on the graph. We use the graph attention mechanism j for prediction. However,
Since it is a semi-supervised classification problem, there is a large amount of unlabeled data in the dataset, and since the AMR domain
is not like the image, NLP domain, there is no large dataset like IMAGENET. Therefore, direct use of the graph attention mechanism
(GAT)\cite{velickovic2017graph} on a small dataset with a large amount of unlabeled data can easily lead to overfitting. For this reason,
We use the label propagation algorithm (LPA)\cite{wang2020unifying} with GAT to perform stacking for prediction.
\begin{figure}[htbp]
	\centering
	\includegraphics[width=\linewidth,scale=1.00]{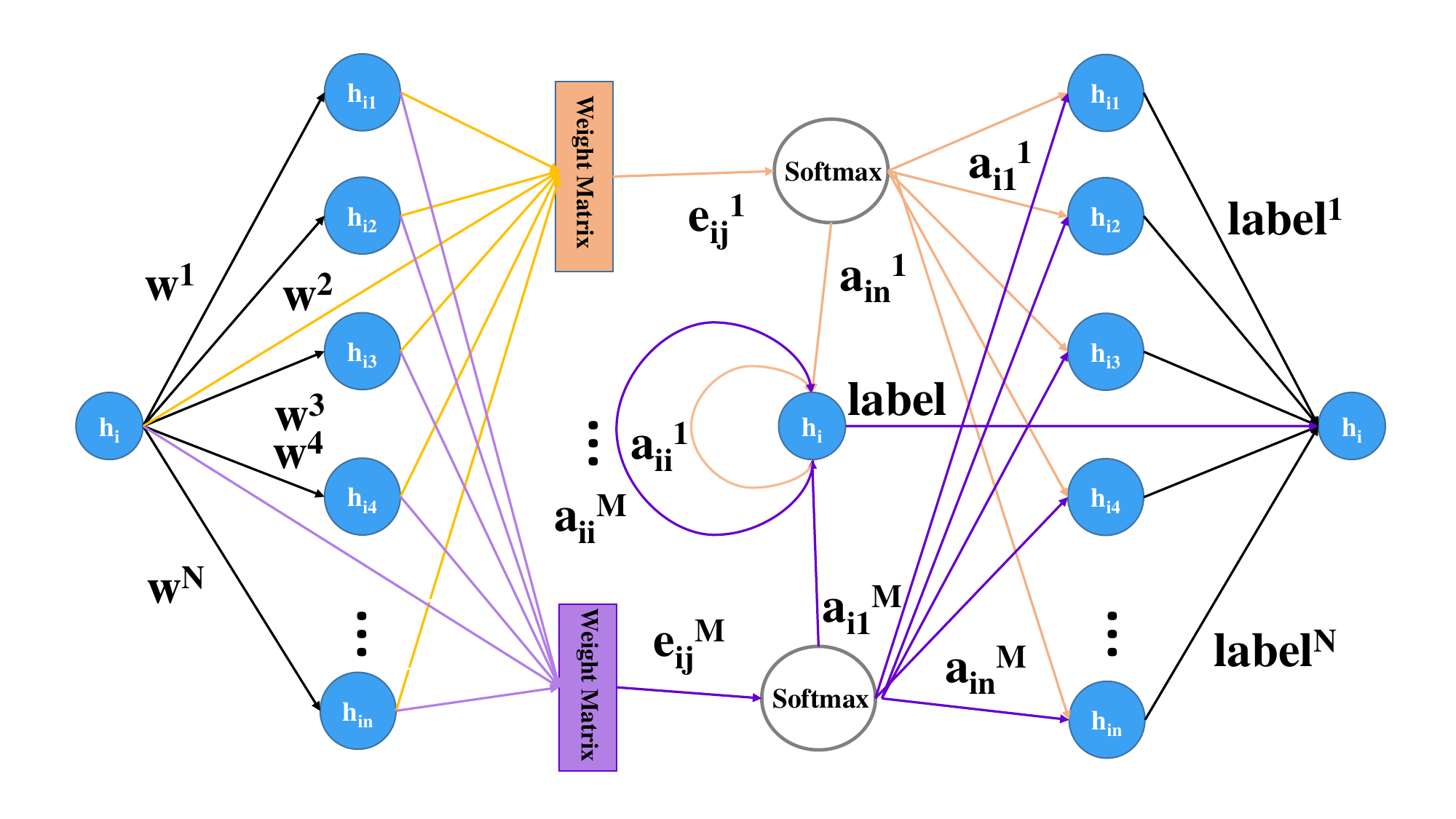}
\caption{GAT-LPA}
	\label{gl}
 \vspace{-0.2cm}
\end{figure}
Suppose a graph has a node $N$ and the dimensional characteristics of the nodes are $L$. Then, the set $X$ can be expressed as:
\begin{equation}
  X=[\overrightarrow{X}_1,\overrightarrow{X}_2,\overrightarrow{X}_3,\dots,\overrightarrow{X}_N],\overrightarrow{X}_i\in R^L
\end{equation}
As illustrated in Fig. \ref{gl}, first, we introduce different functions and computing mechanisms to
calculate the similarity or correlation between the two nodes according to the characteristics of the node itself and the
characteristics of its neighbors. The most common algorithms include finding the vector dot product of the two and finding
the vector similarity of the two properties or by reintroducing additional neural networks for evaluation. In this paper,
The correlation between nodes is obtained by directly calculating the dot product of the feature vectors of two nodes.
The dimension of the feature vector may change during this process: $F->F^{\prime}$. During this process, the dimension
of the feature vector may change. To retain sufficient expressive power, we first need to convert the input
features into high-order features. We initialize a metric of learnable parameters and perform linear transformation
with node features.
\begin{equation}
  \begin{split}
      e_{ij}=LeakyReLU(f(w\overrightarrow{X}_1,w\overrightarrow{X}_2))\\
      w \in R^{L \cdot L^{\prime}}\\
  \end{split}
\end{equation}
where $w$ is a linear transformation of a learnable parameter, $f$ is the vector dot product, $\overrightarrow{X}_i,\overrightarrow{X}_j$
represents any two nodes with connected edges, LeakyReLU makes the whole transformation process have a certain nonlinearity. we consider
the relevance of all neighboring nodes in the target node's neighborhood to the target node (including its own influence) to
better distribute weights among different nodes. Afterward, a calculation algorithm similar to normalization is introduced
to convert the first stage score to a value between 0 and 1. On the one hand, a normalization process can be performed to
sort the original calculated scores into a probability distribution where the sum of all element weights is 1. On the
other hand, through the internal mechanism of the normalization function, the weights of important elements can also be
more emphasized
{\small
\begin{equation}
  a_{ij}=\frac{exp(LeakyReLU(f(w\overrightarrow{X}_i,w\overrightarrow{X}_k)))}{\Sigma_{k \in N_i}exp(LeakyReLU(f(w\overrightarrow{X}_i,w\overrightarrow{X}_k)))}
\end{equation}
}
where $a_{ij}$ is the normalized coordinate system and $N_i$ is the set of neighbors of the node. Finally, we take the normalized
result as the weight coefficient corresponding to the feature and aggregate the neighboring nodes and their own features by weighting
and summing all the neighboring nodes of the current node.
\begin{equation}
    \overrightarrow{X^{\prime}}_i=\sigma(\Sigma_{k \in N_i}a_{ik}W\overrightarrow{X}_i)
\end{equation}
$\sigma$ is a nonlinear activation function. This formula implies that the output features of a node are related to all its neighboring
nodes, represented by a nonlinear transformation of its node features aggregated with those of its neighbors. The nonlinear transformations
and learnable weight parameters are optimized by back propagation of the neural network work.
We repeat the above process for another $m-1$ rounds, thereby having $m$ embedding vectors $\overrightarrow{X^{\prime \lambda}_i}$ $(\lambda\in[1,m])$ for node $n_i$.
In what follows, we calculate the average of the $m$ embedding vectors for each node $n_i$ and feed the result into an activation function $\sigma(\cdot)$. The result $\overrightarrow{X^{\prime}_i}$ is referred to as the node aggregation feature of $n_i$.
\begin{equation}
    \overrightarrow{X^{\prime}_i}=\sigma\Big(\frac{1}{m} \Sigma_{\lambda=1}^{m}\overrightarrow{X^{\prime \lambda}_i}\Big)
\end{equation}
On the other hand, we use the following equation to calculate the weight of each edge in the graph.
\begin{equation}
    E'_{i,j}=\frac{1}{m} \Sigma_{\lambda=1}^{m}a_{i,j}^{\lambda}
\end{equation}
Then, we use $\overrightarrow{X^{\prime}_i}$ and $E'_{i,j}$ to update the node embedding vectors and edge weights of graph $\mathcal{G}$.
In the GAT algorithm process, unlabeled data are subjected to feature aggregation even if they are not involved.
in the final training and prediction phase. Compared with traditional machine learning algorithms, the GAT algorithm can utilize
unlabeled data more effectively.

We use the weight matrix E obtained in the GAT process as the learnable probability transfer matrix P in label propagation. Afterward
We construct the label matrix F for all the data. Assuming that there are $C$ categories and $L$ labeled samples, we define a $LxC$ label
matrix $Y_L$ with the ith row denoting the one-hot encoding vector after label smoothing of the ith sample. Similarly, we also give U unlabeled samples a $UxC$
label matrix $Y_U$, with each row of $Y_U$ having an element of onehot encoding of all 0. Combining them, we obtain a $NxC$ soft label
matrix $F=[YL.YU]$. soft label means that we keep the probability that sample $i$ belongs to each class, not mutually exclusive, and this
sample belongs to only one class with probability 1.

First, we perform propagation, which is the multiplication of the matrix $P$ and the matrix F by $F = PF$.
\begin{equation}
  \begin{aligned}
  &P=\left[\begin{array}{ll}
  P_{L L} & P_{L U} \\
  P_{U L} & P_{U U}
  \end{array}\right]\\
  &f_U \leftarrow P_{U U} f_U+P_{U L} Y_L
  \end{aligned}
\end{equation}
In this step, each node propagates its own label to the other node with a probability determined by P. If two nodes are more similar
(closer in Euclidean space), then the other's label is more likely to be given by its own label. We use the asynchronous update algorithm
to update the label. For the asynchronous update algorithm, the update formula is:
{
\scriptsize
\begin{equation}
  C_x(t)=f\left(C_{x_{i_1}}(t),\cdots,C_{x_{i_m}}(t),\cdots,C_{x_{x_k}}(t-1)\right)
\end{equation}
}
where the labels of neighboring nodes $X_1,...,X_m$ have been updated in the iteration, and then their latest labels are used. The neighboring
nodes $X_{(m+1)},...,X_k$ have not been updated at the time of iteration, then the original labels are used for these neighboring nodes\cite{raghavan2007near}.
The above steps are repeated until convergence (the labels of unlabeled nodes in the graph are not changing) or the maximum number of
iterations threshold is reached (the number of iterations designed in this paper is 20).
\begin{equation}
  f_U=\left(I-P_{U U}\right)^{-1} P_{U L} Y_L
\end{equation}
After label propagation is completed, we can obtain the labels of the unlabeled data $f_U$. We use the cross-entropy function for training.
\begin{equation}
  H(p, q)=-\sum_{i=1}^n p\left(x_i\right) \log \left(f_U\left(x_i\right)\right)
\end{equation}
Since inputting labels in a neural network can lead to label leakage, we borrowed from this paper\cite{shi2020masked} and used a
masking mechanism to train the masked data by applying a random mask to the data during the training process, thus allowing us to train
the data without leaking the self-loop labels.

Finally, we stack the GAT module and the LPA module. specifically, we use the features extracted from the GAT module to
predict the residuals between the predicted results of the LPA module and the true labels. The loss function uses the mean square error
L2 function.
\begin{equation}
  \arg \; \min \; L_2(H(p,q),MLP(GAT(X_i)))
\end{equation}
The loss function can also be considered a regular term of LPA to prevent overfitting. After introducing the basic modules of the
model, we start to build the overall structure of the model. We use the Inception structure\cite{szegedy2015going} to construct four
Inception branches containing four layers of GAT modules and use residual connections\cite{he2016deep} to avoid the over-smoothing problem
in the graph neural network.

\section{Performance Evaluation}\label{IMPLEMENTATION AND EVALUATION}
\begin{figure*}[htbp]
% \vspace{0.1in}
% %\hspace{-15mm}
  \centering
\subfigure[Macro-Average Precision]{
\begin{minipage}[t]{0.24\linewidth}
  \centering
  %\hspace{-15mm}
  \includegraphics[width=0.9\linewidth]{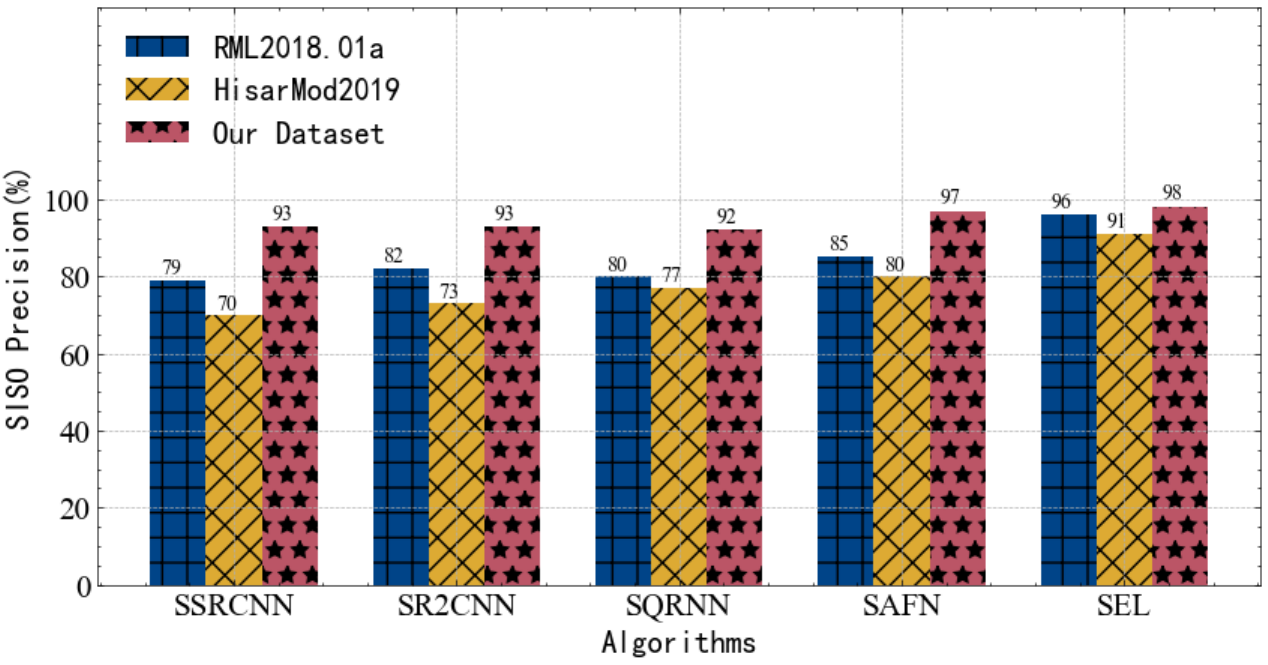}
  % \vspace{5mm} %调整纵向距离
  %\hspace{-15mm}
% \caption{fig1}
\end{minipage}%
}%
% \hspace{-0.3em}
% \hspace{-1mm}
% \hspace{.-15in}
\subfigure[Accuracy in Our Dataset]{
\begin{minipage}[t]{0.24\linewidth}
  \centering
  %\hspace{-15mm}
  \includegraphics[width=0.9\linewidth]{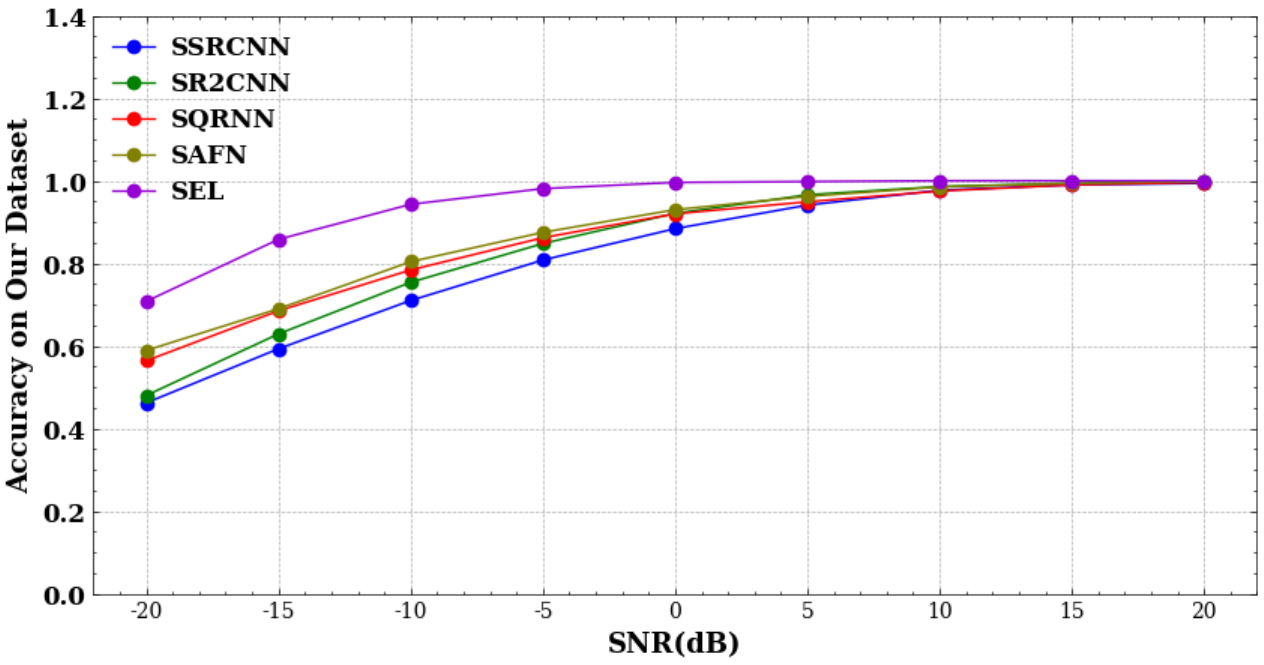}
  %\hspace{-15mm}
% \caption{fig2}
\end{minipage}%
}%
\subfigure[Accuracy in RML2018.01a]{
\begin{minipage}[t]{0.24\linewidth}
  \centering
  %\hspace{-15mm}
  \includegraphics[width=0.9\linewidth]{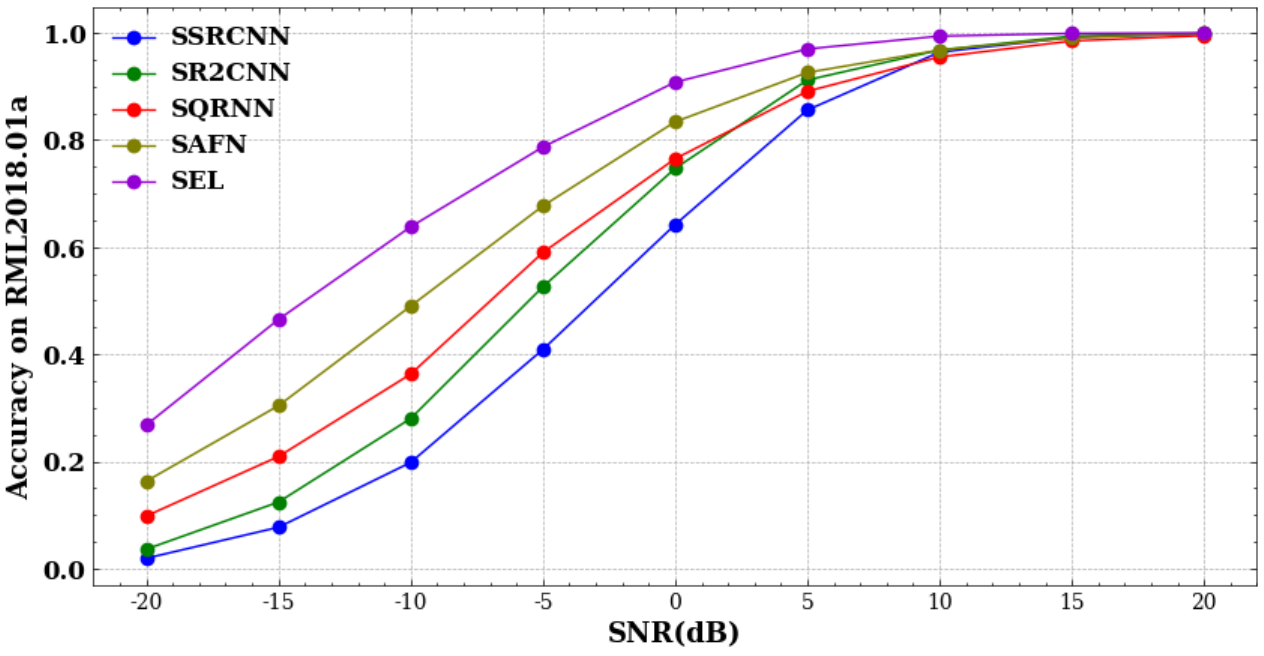}
% \caption{fig3}
%\hspace{-15mm}
\end{minipage}
}%
\subfigure[Accuracy in HisarMod2019.1]{
\begin{minipage}[t]{0.24\linewidth}
  \centering
  %\hspace{-15mm}
  \includegraphics[width=0.9\linewidth]{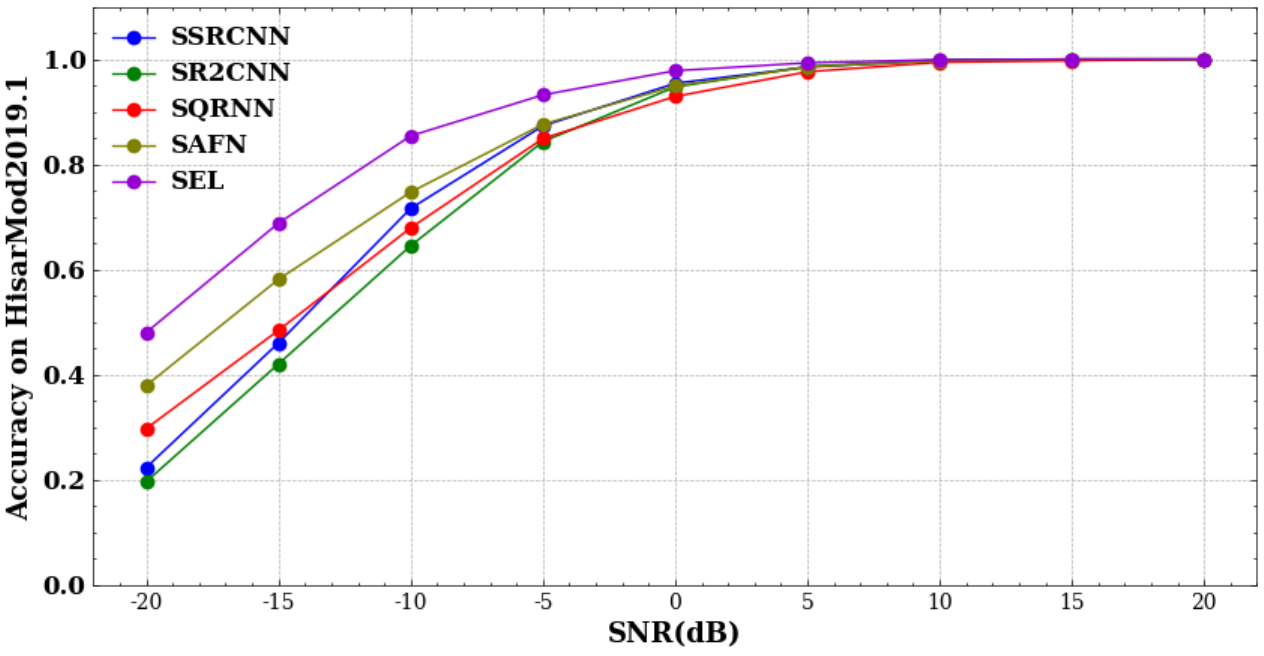}
% \caption{fig4}
%\hspace{-15mm}
\end{minipage}
}%
  \centering
\caption{Accuracy and macroaverage precision of baseline and the algorithms in this paper on SISO scenarios}
  \label{SISO}
\end{figure*}
\begin{figure*}[htbp]
% %\hspace{-15mm}
  \centering
  % %\hspace{-15mm}
\subfigure[Macro-Average Precision]{
\begin{minipage}[t]{0.24\linewidth}
  \centering
  %\hspace{-15mm}
  \includegraphics[width=0.9\linewidth]{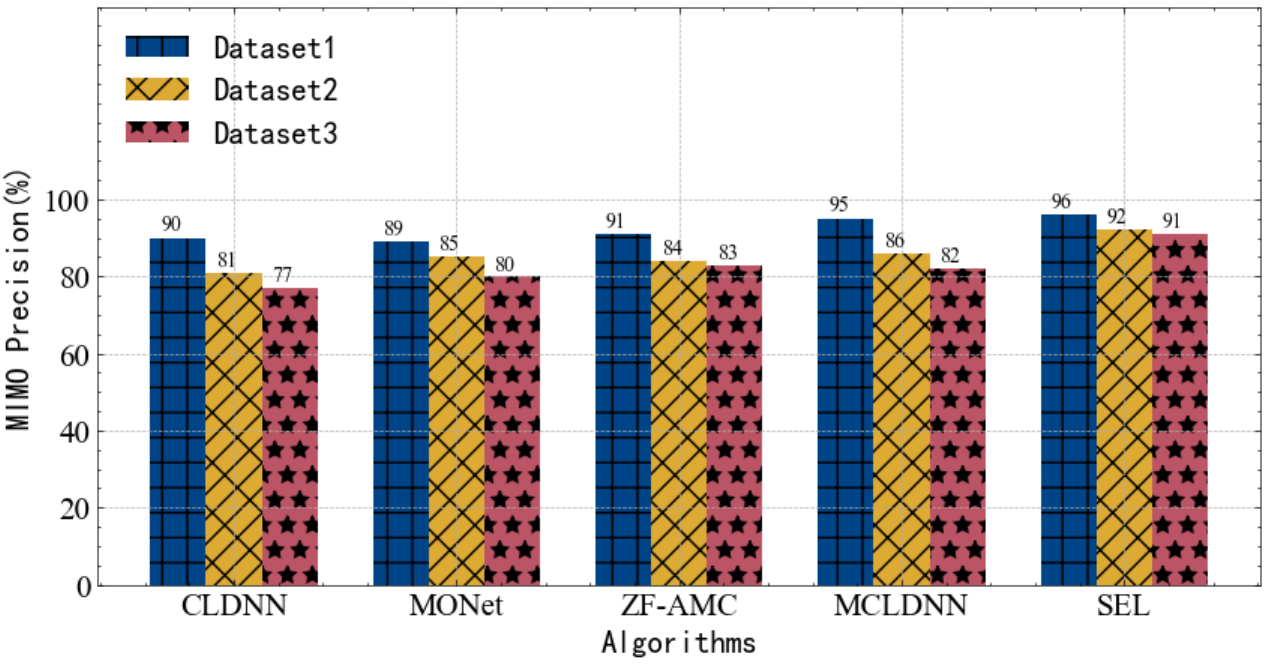}
  %\hspace{-15mm}
% \caption{fig1}
\end{minipage}%
}%
\subfigure[Accuracy in Our Dataset1]{
\begin{minipage}[t]{0.24\linewidth}
% \hspace{-20mm}
  \centering
  %\hspace{-15mm}
  \includegraphics[width=0.9\linewidth]{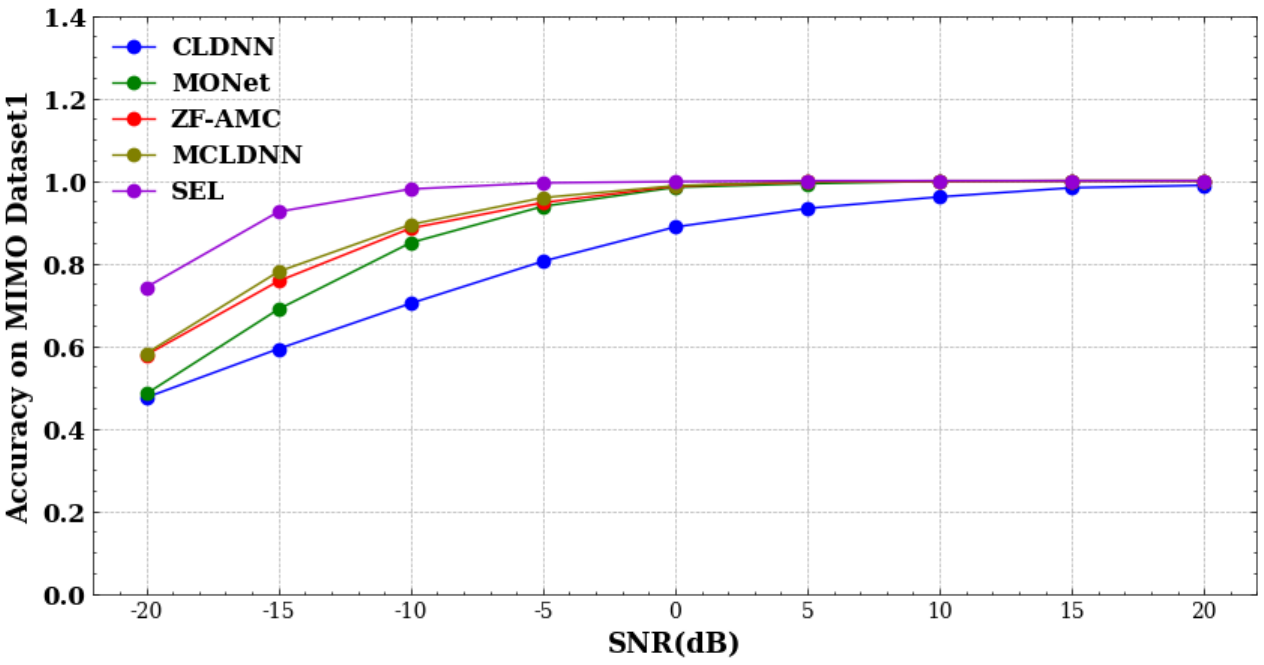}
  %\hspace{-15mm}
% \caption{fig2}
\end{minipage}%
}%
\subfigure[Accuracy in Our Dataset2]{
\begin{minipage}[t]{0.24\linewidth}
  \centering
  %\hspace{-15mm}
  \includegraphics[width=0.9\linewidth]{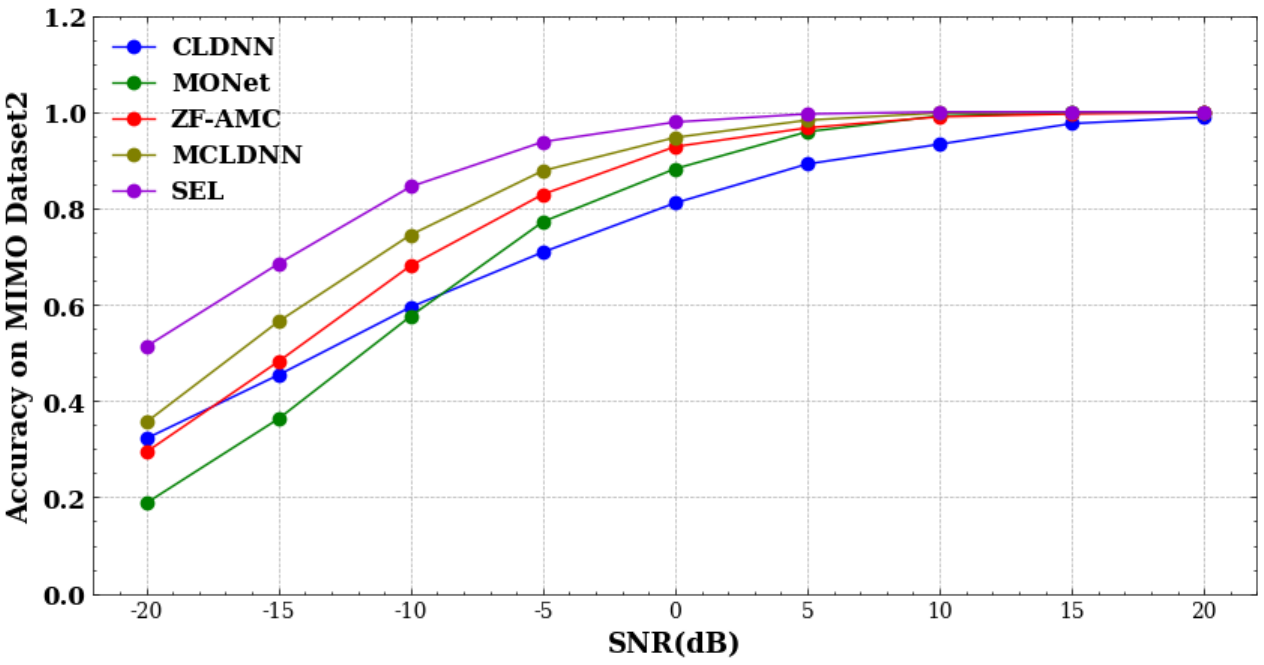}
  %\hspace{-15mm}
% \caption{fig3}
\end{minipage}
}%
\subfigure[Accuracy in Our Dataset3]{
\begin{minipage}[t]{0.24\linewidth}
  \centering
  %\hspace{-15mm}
  \includegraphics[width=0.9\linewidth]{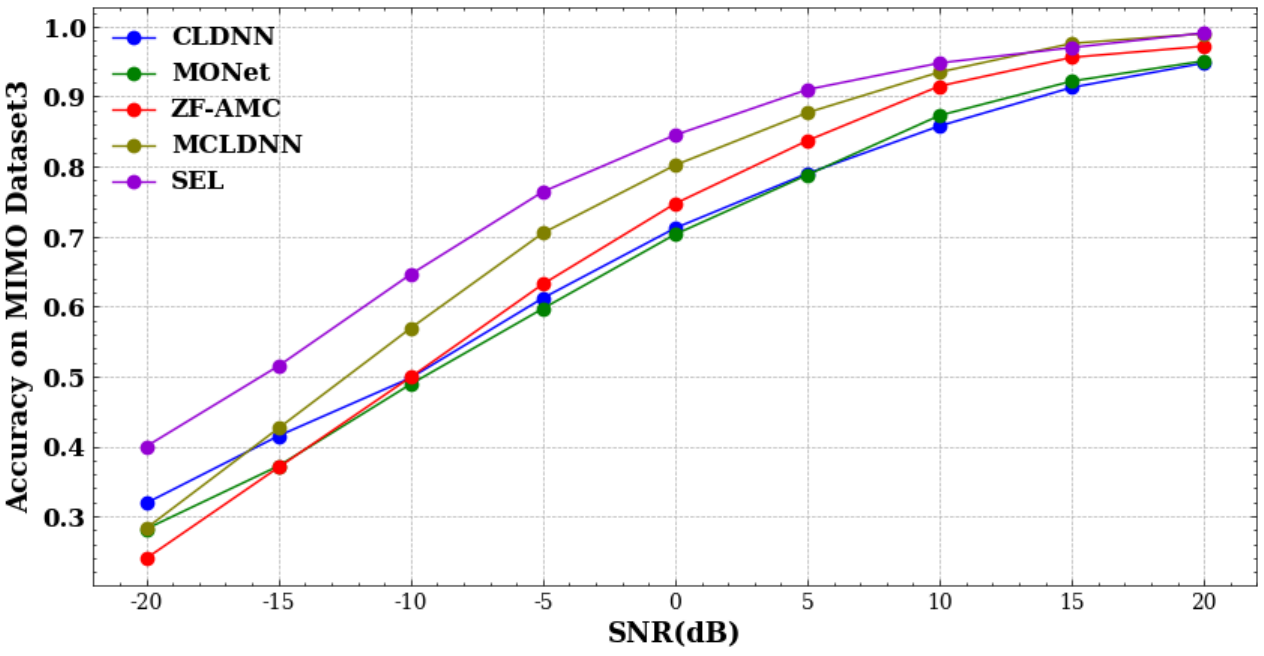}
  %\hspace{-15mm}
% \caption{fig4}
\end{minipage}
}%
  \centering
\caption{Accuracy and macroaverage precision of the baseline and the algorithms in this paper on MIMO scenarios}
  \label{MIMO}
\end{figure*}
In this section, we conduct experiments on several public datasets to evaluate the performance of our algorithm.
\subsection{Dataset and Baseline}
We simulate the real signal through GNU Radio to form the IQ dataset, covering 1,210,000 data samples of various digital and analog modulated
signals modulated with different signal-to-noise ratios, and the current dataset contains 16qam,2ask,2fsk,2psk,32qam,4ask,4fsk,4psk,
64qam,8fsk. 8psk total 11 modulation data, each data signal-to-noise ratio -20 dB-20 dB (no interval), the data are disordered, where
Data column 0 stores the data number, columns 1-1024 store the I-way data, columns 1025-2048 store the Q-way data, and column 2049
stores the modulation label. The dataset contains various signal systems, such as SISO, 4*2 (Nt = 4, Nr = 2), 16*4 (Nt = 16, Nr = 4),
and 64*16 (Nt = 64, Nr = 16) MIMO systems. To verify the generalization of the proposed algorithm in this paper, we use two publicly
available datasets RML2018.01a\cite{o2018over} and HisarMod2019.1\cite{tekbiyik2020robust} with the same modulation as in the training
dataset as the test set to verify the generalization ability of the algorithm in this paper for SISO systems. For the MIMO system, we
simulated two more test datasets because of the small number of public datasets, in which the channel conditions of the first test data
Dataset2 is different from the training data, and the second test dataset Dataset3 contains 3*3 (Nt = 3, Nr = 3), which is a MIMO system.
not seen in the training data. This is used to verify the generalization ability of the model to MIMO systems. We divide our dataset
into a training set, a validation set and a test set at a ratio of 6:2:2. where the training set is further divided into labeled data
and unlabeled data according to the ratio of 1:1.

Since there are no algorithms that can be applied to arbitrary signal systems simultaneously, in this paper, we select several common
data-driven algorithms applied to SISO systems and several common data-driven algorithms applied to MIMO systems as baseline algorithms
for comparison.
% \begin{enumerate}[0]
% \item[-] SISO:

% Among the representative work based on the SISO system are:
The representative SISO based works are as follows.
\begin{enumerate}
\item SSRCNN\cite{dong2021ssrcnn}: proposes a semi-supervised learning (SSL) framework that can efficiently extract knowledge from unlabeled data by designing loss functions and neural network structures.
\item SR2CNN\cite{dong2021sr2cnn}: proposes a CNN-based deep zero-sample learning framework for extracting semantic features while maintaining the performance of the decoder and classifier.
\item SQRNN\cite{ghasemzadeh2022gs}: proposes an automatic constraint classifier architecture that exploits the low time slot feature of the transformation threshold to enhance the learning capability of the model.
\item SAFN\cite{chang2021multitask}: proposes a framework that can explore interaction features and temporal information by fusion.
\end{enumerate}
% \item[-] MIMO:
% Among the representative work based on the MIMO system are:

The representative MIMO based works are as follows.
\begin{enumerate}
\item CLDNN\cite{west2017deep}: use Inception structure and LSTM to improve learning synchronization and equalization.
\item MONet\cite{huynh2022mimo}: uses a 3D convolutional network with combined residual connectivity.
\item ZF-AMC\cite{wang2020automatic}: propose a framework that applies ZF equalization techniques and is suitable for MIMO systems.
\item MCLDNN\cite{xu2020spatiotemporal}: use an effective model structure and extract features from a spatiotemporal perspective.
\end{enumerate}
% \end{enumerate}
%
% \begin{enumerate}[0]
%     \item[$\bullet$] A is greater than B A is greater than B A is greater than B A is greater than B A is greater than B A is greater than B A is greater than B A is greater than B A is greater than B A is greater than B A is greater than B
%     \item[$\bullet$] B is greater than C
%     \item[$\bullet$]  C is greater than D
% \end{enumerate}
%

Since the application scenarios of the baseline algorithm vary in terms of learning, we use the dimensional transformation algorithm to
Adjust the input and output dimensions of the model to fit the input data and keep the same intermediate parameters in all datasets.
We demonstrate the effectiveness of our algorithm by comparing the backbone network of the algorithm proposed in this paper with the
backbone networks of other algorithms.
\subsection{Experimental Setup and Evaluation Metrics}
The hardware components of the experiments include a laptop and a high-performance server with an NVIDIA Tesla P100 GPU Powerleader PR2730G.
For the software components, we use python, PYG, Pandas, Numpy, Networkx, and Pytorch from Matplotlib for data processing, model building,
and visualization.

The algorithm proposed in this paper and other baseline algorithms use the Adam optimizer to run 300 epochs per 64 data points as a batch and
train three times for averaging to obtain the final results.

We use the macroaverage precision and accuracy at different signal-to-noise ratios as the metric in this paper.
\begin{equation}
    \begin{aligned}
        P_{macro}=\frac{1}{k}\sum_{i=1}^{k}\frac{TP_i}{TP_i+FP_i}\\
        Accuracy=\frac{\sum_{i=1}^{k}l_i}{\sum_{i=1}^{k}m_i}\\
    \end{aligned}
\end{equation}
where $i$ denotes the number of samples and $l_i$ denotes the number of samples predicted by the model to be the $i$-th class and actually belong
to the $i$-th class, and $m_i$ denotes the number of samples predicted by the model to be the $i$-th class.
\subsection{Controlled Experiment}
As illustrated in Fig. \ref{SISO}, the algorithm proposed in this paper achieves SOTA results in various datasets.
under the SISO scenario. In particular, on two publicly available datasets, our algorithm still obtains $0.96$ and $0.91$ accuracy
without any transfer learning or other techniques, and the accuracy under different signal-to-noise ratios is much higher than
other algorithms. This proves our previous assertion that using graph structure to relate sample data together through relationships
and then making joint inference through the topology between data has higher robustness and higher accuracy than using data alone.
Meanwhile, in the semi-supervised learning scenario where there is a large amount of unlabeled data in the dataset, the baseline algorithm
cannot effectively utilize those unlabeled data, while our algorithm can use the relationships between data to perform feature aggregation
to smooth out the noise and enhance the generalization ability of the model under different channel conditions.

As illustrated in Fig. \ref{MIMO}, in the MIMO scenario, our algorithm achieves SOTA criteria on all three datasets constructed by
our simulation. It shows that our algorithm has higher robustness and accuracy compared to the baseline algorithm, and the input data
dimension of our algorithm is variable, which allows our algorithm to work in scenarios where MIMO and SISO coexist compared to traditional
machine learning-based AMR algorithms with fixed input dimensions. In the case of dataset2, where the channel conditions are significantly
different from the training set, our algorithm can extract more robust features by aggregating features to fit the effects of different
channel conditions using the relationship between the data compared to the baseline algorithm. For dataset3, which has a 3*3 MIMO that
is not visible in the training set, the traditional algorithm can only transform the input dimension to the same dimension by interpolating
the complementary or discarding algorithm using dimensional transformation. This leads to redundancy or loss of features, resulting in reduced
model prediction accuracy, while the MIMO embedding in this algorithm is generalizable than traditional algorithms because it uses
an inductive graph inference algorithm that can be applied to dimensionally variable and order-independent data.

% 为了验证在发射天线数量未知的情况下使用可训练张量作为节点特征的必要性，我们对节点特征进行了随机初始化实验。结果表明，与使用可训练张量相比，随机初始化节点特征并不会影响最终的实验精度。关键在于图网络的构建和训练，无论初始节点设置如何，图网络都能学习到有效的特征。随机初始化为未知的天线条件提供了更方便、更灵活的方法。
% 总之，实验结果表明，对于未知的发射天线情况，节点特征的随机初始化是一种可行而有效的技术，既简化了图的构建，又保持了预测的准确性。
In order to verify the necessity of using trainable tensor as node characteristics under unknown number of transmit antennas, we conduct experiments using random initialization for node features.The results show that randomly initializing the node features does not affect the final experimental accuracy compared with using trainable tensors. This suggests that the initial node features are not the determining factor for performance. The key lies in the construction and training of the graph networks, which can learn effective features regardless of the initial node settings. 
Moreover, random initialization provides a more convenient, flexible and scalable approach for unknown or expanding antenna conditions, where the trainable tensor needs to be re-configured whenever new nodes are introduced. 
The random initialization simplifies the graph construction procedure without compromising predictive power.
In conclusion, the experimental results demonstrate that random initialization of node features is a feasible and effective technique for unknown transmit antenna scenarios, simplifying the graph construction while maintaining prediction accuracy.
% \vspace{-0.5cm}
\subsection{Ablation Experiment}
 % \vspace{-0.1cm}
\begin{table}[htbp]
  \centering
  \caption{Results of ablation experiments}
  \label{Ae}
  \begin{tabular}{|c|cll|cll|}
    \hline
    \begin{tabular}[c]{@{}c@{}}Macro-average precision\\ of our dataset\end{tabular} & \multicolumn{3}{c|}{SISO} & \multicolumn{3}{c|}{MIMO} \\ \hline
    \begin{tabular}[c]{@{}c@{}}Dimensional transformations\\ MIMO embedding\end{tabular}                      & \multicolumn{3}{c|}{0.94} & \multicolumn{3}{c|}{0.92} \\ \hline
    \begin{tabular}[c]{@{}c@{}}Complete graph\\ KNN graph\end{tabular}                           & \multicolumn{3}{c|}{0.93} & \multicolumn{3}{c|}{0.90} \\ \hline
    \begin{tabular}[c]{@{}c@{}}GAT\\ GAT-LPA\end{tabular}                             & \multicolumn{3}{c|}{0.96} & \multicolumn{3}{c|}{0.91} \\ \hline
    \begin{tabular}[c]{@{}c@{}}SEL\\ No changes\end{tabular}                             & \multicolumn{3}{c|}{0.98} & \multicolumn{3}{c|}{0.96} \\ \hline
    \end{tabular}
\end{table}
To verify the effectiveness of the modules in the algorithm proposed in this paper, we conducted ablation experiments with
macroaverage precision on a test set divided on the dataset constructed by our simulation. As shown in Table \ref{Ae}, we first replace
our MIMO embedding module with a general dimensional change, using interpolation complementation or discarding to map the input dimension
to a fixed dimension. It can be seen that the accuracy decreases dramatically, even less than some baseline algorithms. This can verify
Our previous inference that forcing features to the same latitude by interpolation or discard will result in feature loss or redundancy
and thus affect the prediction accuracy of the model, while our proposed MIMO embedding algorithm based on subgraph embedding can adaptively
build a mapping from non-Euclidean space to Euclidean space to ensure the integrity of features to the maximum extent. Second, we directly
construct the sample graph as a complete graph to replace the KNN graph building process, and we can see that the accuracy decreases
compared to the original one, and we find that the node features output from the graph neural network become increasingly similar to
the increase in training times. This indicates that directly constructing the topology as a complete graph will more easily lead to
over-smoothing of the graph neural network. Finally, we use GAT directly instead of the GAT-LPA module in the algorithm of this paper,
and we can see that the accuracy also decreases to a certain extent and the convergence of the model becomes slower. This shows that
Using the LPA mechanism to assist GAT for semi-supervised learning has better results and faster convergence than using GAT alone.

\section{Related Work}\label{RELATED WORK}
% \subfile{section/rela.tex}
Existing related work can be categorized into signal modulation identification based on SISO systems and signal modulation identification based on specific MIMO systems.
\subsection{DL Models for AMR in SISO Systems}
Lee \emph{et al.}\cite{lee2017deep} constructed a four-layer fully connected network to perform signal modulation recognition.
liu \emph{et al.}\cite{liu2017deep} used CNN to extract signal features.
Zhang \emph{et al.}\cite{zhang2021real} used the DenseNet structure and residual connections to construct a deep neural network for signal modulation recognition.
Liu \emph{et al.}\cite{liu2020modulation} used CNN networks to extract the features of the signals, followed by similarity to construct the adjacency matrix between the signal samples and use graph convolution for signal modulation recognition.
Some papers\cite{liu2020modulation,ghasemzadeh2022robust,tonchev2022automatic,ghasemzadeh2023ggcnn} improve the recognition accuracy by converting the modulated signal segmentation into graph form and feeding it into a graph convolutional network.
SSRCNN\cite{dong2021ssrcnn} proposes a semi-supervised learning (SSL) framework that can efficiently extract knowledge from unlabeled data by designing loss functions and neural network structures.
SQRNN\cite{ghasemzadeh2022gs} proposed an automatic constraint classifier architecture that exploits the low time slot feature of the transformation threshold to enhance the learning capability of the model.
However, none of the above algorithms consider scenarios where there is a large amount of unlabeled data and multiple signal systems. As a result, some key information is lost, and they cannot represent the target problem comprehensively.
\subsection{DL Models for AMR in MIMO Systems}
CLDNN\cite{west2017deep} uses Inception structure and LSTM to improve learning synchronization and equalization.
MCLDNN\cite{xu2020spatiotemporal} uses an effective model structure and extracts features from a spatiotemporal perspective.
Qi \emph{et al.}\cite{qi2020automatic} used multimodal deep learning using CNN to extract features of IQ data and channel state information separately and later used feature fusion to increase the accuracy of the model.
Hou \emph{et al.}\cite{hong2017automatic} used the GRU model with RNN architecture to predict the class of signal modulation.
MONet\cite{huynh2022mimo} uses a 3D convolutional network with combined residual connectivity.
ZF-AMC\cite{wang2020automatic} proposes a framework that applies ZF equalization techniques and is suitable for MIMO systems.
Xiong \emph{et al.}\cite{xiong2021mathtt} used self-similarity to implement automatic signal modulation identification in large-scale MIMO systems.
All the above algorithms are only applicable to specific channel states and specific signal systems and do not consider the complexity in real scenarios. hence, the generalization capability is low.
\section{Conclusion}\label{CONCLUSION}
We proposed a algorithm for electromagnetic signal modulation recognition based on subgraph embedding, in which the signal samples are viewed as a large graph, and the signal system corresponding to each sample data is considered as a subgraph. 
We utilized the concept of subgraph embedding to extract the features of subgraphs using graph isomorphic networks and constructed a mapping from graph space to Euclidean space using set2set. 
Afterwards, we used the KNN graph algorithm to construct the topology between sample data, employed the relationship between samples to smooth the effects of noise and different channel conditions to extract features with high robustness, and used the label propagation algorithm for predictions on a semi-supervised problem with a large amount of unlabeled data. 
Experimental results show that the algorithm proposed in this paper has higher accuracy and robustness then the existing state-of-the-art algorithms.
% \cleardoublepage
\bibliographystyle{IEEEtran}
\bibliography{IEEEexample}
\end{CJK}
\end{document}